\documentstyle[12pt]{article}
\begin{document}
\def\bea{\begin{eqnarray}}
\def\eea{\end{eqnarray}}
\title{\bf {Cosmological Particle Creation and Dynamical Casimir Effect}}
\author{
M.R. Setare  \footnote{E-mail: rezakord@yahoo.com}
\\
{Department of Science, Physics group, Kurdistan University,
Sanandeg, Iran }\\
{  Department of Physics, Sharif University of Technology, Tehran,
Iran }\\{Institute for Theoretical Physics and Mathematics,
Tehran, Iran }}

\date{\small{\today}}
 \maketitle
\begin{abstract}
In this paper we have considered the particle creation in the
spatially closed Robertson-Walker space-time. We considered a real
massive scalar field which conformally coupled to the
Robertson-Walker background. With the dependence of the scale
factor on time, the case under consideration is a dynamical
Casimir effect with moving boundaries.

 \end{abstract}

\newpage

 \section{Introduction}
   The Casimir effect is one of the most interesting manifestations
  of nontrivial properties of the vacuum state in a quantum field
  theory (for reviews see \cite{mueller,Moste,Milt,Birrell}) and
   can be viewed
  as a polarization of
  vacuum by boundary conditions. A new phenomenon, a
  quantum creation of particles (the dynamical Casimir effect)
  occurs when the geometry of the system varies in time. In two
  dimensional spacetime and for conformally invariant fields the
  problem with dynamical boundaries can be mapped to the
  corresponding static problem and hence allows a complete study
  (see \cite{Moste,Birrell} and references therein). In higher
  dimensions the problem is much more complicated and is solved
  for some simple geometries. The vacuum stress induced by uniform
  acceleration of a perfectly reflecting plane is considered in
  \cite{Cand}. The corresponding problem for a sphere
  expanding in the four-dimensional spacetime with constant
  acceleration is investigated by Frolov and Serebriany
  \cite{Frol1,Frol2} in the perfectly reflecting case and by Frolov
  and Singh \cite{Frol3} for semi-transparent boundaries. For more
  general cases of motion by vibrating cavities the problem of particle and
  energy creation is considered on the base of various
  perturbation methods \cite{Caluc, Hacyan,Widom,dod,Lamb,Ji,Schut,dod1}(for more
  complete list of references see \cite{dod1}).
   It have been
  shown that a gradual accumulation of small changes in the quantum
  state of the field could result in a significant observable
  effect. A new application of the dynamical Casimir effect has
  recently appeared in connection with the suggestion by Schwinger
  \cite{Schwing} that the photon production associated with changes in
  the quantum electrodynamic vacuum state arising from a
  collapsing dielectric bubble could be relevant for
  sonoluminescence (the phenomenon of light
  emission by a sound-driven gas bubble in a fluid \cite{Bar}).
  For the further developments and discussions this quantum-vacuum
  approach see \cite{Milton1,Eber,lib,MilNg,Liber2} and references
  therein.\\
  The possibility of particle production due to space-time
  curvature has been discussed by Schrodinger \cite{sch}, while
  other early work is due to DeWitt \cite{de}, and Imamura
  \cite{im}. The first thorough treatment of particle production
  by an external gravitational filed was given by Parker
  \cite{{pa1},{pa2}}. Particle creation from the quantum scalar
  vacuum by expanding or contracting spherical shell with
  Dirichlet boundary conditions is considered in \cite{set}. In
  another paper the case is considered when the sphere radius
  performs oscillation with a small amplitude and the expression
  are derived for the number of created particles to the first
  order of the perturbation theory \cite{set1}. Now in the present
  paper by using the result of \cite{set} we consider particle
  creation in closed Robertson-Walker space-time, when the scale
  factor represent an asymptotically static space-time.

\section{Gravitational particle creation}
In flat space-time, Lorentz invariance is a guide which generally
allows to identify a unique vacuum state for the theory. However,
in curved space-time, we do not have Lorentz symmetry. In general,
there does not exist a unique vacuum state in a curved space-time.
As a result, the concept of particles becomes ambiguous, and the
problem of the physical interpretation becomes much more difficult
\cite{{Milton1},{for}}. The particle creation by an expanding
universe was first hinted at in the work of Schrodinger
\cite{sch}, this phenomenon first carefully investigated by Parker
\cite{{pa2},{pa3}}. We restrict our attention to the case of
spatially closed Robertson-Walker universe which metric is as
following

\begin{equation} \label{metric1}
ds^{2}=a^{2}(\eta)(d\eta^{2}-dl^{2}),
\end{equation}

\begin{equation}\label{metric2}
dl^{2}=d\chi^{2}+\sin^{2}\chi(d\theta^{2}+\sin^{2}\theta
d\varphi^{2}).
\end{equation}
where $a(\eta)$ is the scale factor and $\eta$ is conformal time,
$0 \leq \chi \leq \pi$.
 Let us consider a real massive scalar field which coupled to the
 closed Robertson-Walker background. With the dependence of the
 radius of curvature $ a(\eta)$ on time, the case under
 consideration is a dynamical Casimir effect with a moving
 boundaries \cite{set}.
 The corresponding wave equation is

\begin{eqnarray} \label{waeq}
(\Box + m^{2} +\xi R)\phi=0,
\end{eqnarray}
where $R$ is the scalar curvature
\begin{equation}\label{cu}
R=\frac{6(a''+a)}{a^{3}},
\end{equation}
where prime stands for the conformal time-derivative, $\xi$ is a
coupling constant, here we consider the conformal coupling
$\xi=1/6$, in this case the (\ref{waeq}) as \cite{bor}
\begin{equation}\label{fieq}
\phi''(x)+\frac{2a'}{a}\phi'(x)-\Delta^{(3)}\phi(x)+(m^{2}a^{2}+\frac{a''}{a}+1)\phi(x)=0,
\end{equation}
where $\Delta^{(3)}$ is the angular part of the Laplacian operator
on a 3-sphere. The solutions of (\ref{fieq}) are
\begin{equation}\label{solu}
\phi^{(+)}_{\lambda l
M}(x)=\frac{1}{\sqrt{2}a(\eta)}g_{\lambda}(\eta) \phi
{\star}_{\lambda l M}(\chi, \theta, \varphi).
\end{equation}
The eigenfunctions  of the three-dimensional Laplasian are as
\begin{equation}\label{eig}
 \phi_{\lambda l M}(\chi, \theta, \varphi)=\frac{1}{\sqrt{\sin \chi}}\sqrt{\frac
 {\lambda (\lambda +l)!}{(\lambda
 -l+1)!}}P_{\lambda-1/2}^{-l-1/2}(\cos \chi)Y_{lM}(\theta
 ,\varphi),
 \end{equation}
$\lambda =1,2,..., L=0,1,2,...,\lambda-1$ ,$Y_{l,M}$ are spherical
harmonics, and $P_{\mu}^{\nu}(z)$ are the adjoint Legendre
functions on the cut. The time-dependent function $g_{\lambda}$
satisfies the oscilatory equation \cite{bor}
\begin{equation}\label{osc}
g''_{\lambda}(\eta) +
\omega^{2}_{\lambda}(\eta)g_{\lambda}(\eta)=0,
\end{equation}
where
\begin{equation}\label{freq}
\omega_{\lambda}^{2}(\eta)=\lambda^{2}+m^{2}a^{2}(\eta).
\end{equation}
Let us consider an exactly solvable case when
\begin{equation}\label{aeq}
a(\eta)=\sqrt{A+B\tanh \frac{\eta}{\eta_{0}}} \hspace{2cm} A > B,
\end{equation}
where $A,B$ and $\eta_{0}$ are constants, this is corresponds to
the contraction for $ B<0$ and expansion for $B>0$. The
corresponding frequencies are
\begin{equation}\label{freqth}
\omega_{\lambda}^{2}(\eta)=\lambda^{2}+m^{2}(A+B \tanh
\frac{\eta}{\eta_{0}}).
\end{equation}
For asymptotically static situation at past and future the in-
and out- vacuum states can be defined, where we use the notations
\begin{equation} \label{omegalf}
\omega_{\lambda }^{{\rm in}}=\sqrt{\lambda^{2}+m^{2}a_{-}},
\qquad \omega_{\lambda }^{{\rm
out}}=\sqrt{\lambda^{2}+m^{2}a_{+}}, \qquad a_{\pm}=\lim_{\eta
\rightarrow \pm \infty}a(\eta)
\end{equation}
for the corresponding eigenfrequencies. Now we need to solve the
equation (\ref{osc}) with $\omega_{\lambda}(\eta)$ given by
(\ref{freq}). The corresponding solutions are given by
hypergeometric function. The normalized in- and out- modes are
given by formula \cite{Birrell}
 \begin{eqnarray} \label{hypgeom}
g^{s}_{\lambda}(\eta)&=&(2\omega^{s}_{\lambda})^{-1/2}\exp[-\imath\omega_{\lambda}^{+}\eta-
\imath\omega_{\lambda}^{-}\eta_{0} \ln[2\cosh(\eta/\eta_{0})]] \times \nonumber \\
&\times &  \, {}_{2}F_{1}(1+\imath\omega_{\lambda}^{-}\eta_{0},
\imath\omega_{\lambda}^{-}\eta_{0};1\mp
\imath\omega_{\lambda}^{s}\eta_{0};\frac{1}{2}(1\pm\tanh(\eta/\eta_{0}))),\quad
s={\rm in},\, {\rm out} ,
\end{eqnarray}
where uper/lower sign corresponds to the in/out- modes, and
\begin{equation} \label{omegpm}
\omega ^{\pm}_{\lambda}=\frac{1}{2}(\omega_{\lambda}^{{\rm
out}}\pm \omega_{\lambda}^{{\rm in}}).
\end{equation}
 The corresponding eigenfunctions are related by the Bogoliubov
transformation
 \begin{equation}\label{qinBogdiag}
g^{{\rm (in)}}_{\lambda}=\alpha_{\lambda} g^{{\rm
(out)}}_{\lambda}+\beta_{\lambda} g^{{\rm (out)} *}_{\lambda},
\end{equation}
where $\alpha_{\lambda}$ and $\beta_{\lambda}$ are the Bogoliubov
coefficients. Using the linear relation between hypergeometic
functions, similar to \cite{Birrell} for the coefficients in this
formula one finds
\begin{equation} \label{Bogalf3}
\alpha_{\lambda}=\left( \frac{\omega_{\lambda}^{{\rm
out}}}{\omega_{\lambda}^{{\rm in}}}\right) ^{1/2}
\frac{\Gamma(1-\imath\omega_{\lambda}^{{\rm in}}\eta_{0})
\Gamma(-\imath\omega_{\lambda}^{{\rm out}}\eta_{0})}
{\Gamma(-\imath\omega_{}^{+}\eta_{0})\lambda
\Gamma(1-\imath\omega_{\lambda}^{+}\eta_{0}))},
\end{equation}

\begin{equation} \label{Bogbet3}
\beta_{\lambda}=\left( \frac{\omega_{\lambda}^{{\rm
out}}}{\omega_{\lambda}^{{\rm in}}}\right) ^{1/2}
\frac{\Gamma(1-\imath\omega_{\lambda}^{{\rm
in}}\eta_{0})\Gamma(\imath\omega_{\lambda}^{{\rm out}}\eta_{0})}
{\Gamma(\imath\omega_{\lambda}^{-}\eta_{0})\Gamma(1+
\imath\omega_{\lambda}^{-}\eta_{0})}.
\end{equation}
The mean number of particles produced through the modulation of
the single scalar mode is
\begin{equation} \label{Nex1}
<{\rm in}|N_{\lambda}|{\rm
in}>=|\beta_{\lambda}|^{2}=\frac{\sinh^{2}(\pi\omega_{\lambda}^{-}\eta_{0})}
{\sinh(\pi\omega_{\lambda}^{{\rm
in}}\eta_{0})\sinh(\pi\omega_{\lambda}^{{\rm out}}\eta_{0})}.
\end{equation}
The total number of particles produced is obtained by taking the
sum over all the oscillation modes :
\begin{equation} \label{Ntotex1}
<{\rm in}|N|{\rm in}>= \sum_{\lambda=1}^{\infty}
\frac{\sinh^{2}[\pi
\eta_{0}(\sqrt{\lambda^{2}+(A+B)m^{2}}-\sqrt{\lambda^{2}+(A-B)m^{2}})/2]}{\sinh(\pi
\eta_{0}\sqrt{\lambda^{2}+(A+B)m^{2}})\sinh(\pi
\eta_{0}\sqrt{\lambda^{2}+(A-B)m^{2}})} .
\end{equation}
Therefore the energy related to the particles production is given
by
\bea
 \label{enex1} E&=&\sum_{\lambda=1}^{\infty}
N_{\lambda}\omega_{\lambda}^{{\rm out}}\\
 &=&
\sum_{\lambda=1}^{\infty} \frac{\sinh^{2}[\pi
\eta_{0}(\sqrt{\lambda^{2}+(A+B)m^{2}}-\sqrt{\lambda^{2}+(A-B)m^{2}}
)/2]}{\sinh(\pi \eta_{0}\sqrt{\lambda^{2}+(A+B)m^{2}} )\sinh(\pi
\eta_{0}\sqrt{\lambda^{2}+(A-B)m^{2}} )} \sqrt{\lambda^{2}+
m^{2}(A+B)}.\nonumber \eea
\section{Conclusion}
The creation of particles from the vacuum takes place due to the
  interaction with dynamical external constraints. For example
  the motion of a single reflecting boundary (mirror) can create
  particles \cite{Birrell}, the creation of particles by
  time-dependent external gravitational field is another example of
  dynamical external constraints.\\
   It has been shown
  \cite{{Nugayev1},{Nugayev2}} that particle creation by black
  hole in four dimension is as a consequence of the Casimir effect
  for spherical shell. It has been shown that the only existence
  of the horizon and of the barrier in the effective potential is
  sufficient to compel the black hole to emit black-body radiation with
  temperature that exactly coincides with the standard result for
  Hawking radiation.
In this paper we have considered the particle creation in the
spatially closed Robertson-Walker space-time. We considered a real
massive scalar field which conformally coupled to the
Robertson-Walker background. With the dependence of the scale
factor on time, the case under consideration is a dynamical
Casimir effect. When scale factor represent an asymptotically
static space-time at past and future, the in- and out- vacuum
states can be defined. Then we obtained the Bogoliubov
coefficients, after that the number of particles produced and the
energy related to those can be explicitly found.

\end{document}